\documentclass[12pt]{article}
\pdfoutput=1
\usepackage{jheppub}

\usepackage{amssymb}
\usepackage{color}
\usepackage{hyperref}
\usepackage{amsthm}
\usepackage{graphicx}
\usepackage{dcolumn}
\usepackage{bm}
\usepackage{amssymb}
\usepackage{verbatim}
\usepackage{amscd}
\usepackage{amssymb}
\usepackage{amsmath, amsfonts}
\usepackage{setspace}
\usepackage{amsthm}
\usepackage{enumerate}
\usepackage{subfigure}
\usepackage{tikz}
\usetikzlibrary{decorations.pathreplacing,shapes.misc}
\usepackage{comment}

\newcommand{\bra}[1]{\left\langle #1 \right|}
\newcommand{\ket}[1]{\left|#1\right\rangle}

\newcommand{\Tr}{\textrm{Tr}}

\theoremstyle{definition}

\begin{document}
\title{Precursor problem and  holographic mutual information}
\author[1]{Ning Bao,}
\affiliation[1]{Walter Burke Institute of Theoretical Physics, Caltech, Pasadena, CA}
\affiliation[1]{Institute for Quantum Information and Mathematics, Caltech, Pasadena, CA}

\author[2]{Isaac H. Kim}
\affiliation[2]{Perimeter Institute for Theoretical Physics, Waterloo ON N2L 2Y5, Canada}
\affiliation[2]{Institute for Quantum Computing, University of Waterloo, Waterloo ON N2L 3G1, Canada}
\date{\today}

\abstract{
The recent proposal of Almheiri et al.\cite{Almheiri2014}, together with the Ryu-Takayanagi formula, implies the entanglement wedge hypothesis for certain choices of boundary subregions. This fact is derived in the pure AdS space. A similar conclusion holds in the presence of quantum corrections, but in a more restricted domain of applicability. We also comment on \cite{Dong2016} and some similarities and differences with this work.}
\maketitle

\section{Introduction}
In the context of the AdS/CFT correspondence\cite{Maldacena1998}, there is an ongoing debate on whether bulk operators can be reconstructed from certain subregions of the boundary CFT. There are several natural conditions that follow from causality constraints, but they are insufficient to completely determine the solution to this problem\cite{Bousso2012,Czech2012,Hubeny2012,Almheiri2014,Jafferis2014}. There are toy models for which the problem can be studied more explicitly\cite{Pastawski2015,Yang2015,Hayden2016}, but whether the insights obtained within these models can be applied more generally remains unclear.

Hamilton et al. provided a prescription\cite{Hamilton2006} -- known as the AdS/Rindler reconstruction -- to reconstruct bulk local operators from the CFT;  operators can be reconstructed from a boundary subregion up to $1/N$ corrections if they lie in the causal wedge, i.e., a set of bulk points that can communicate back and forth with the given boundary region. Building upon a work of Morrison\cite{Morrison2014}, the nature of this reconstruction was recently elucidated by Almheiri, Dong, and Harlow\cite{Almheiri2014}. By identifying several puzzles that arise from this prescription, they concluded that the AdS/Rindler reconstruction should not be viewed as a statement about operators, but rather as a statement about operators that are restricted to some subspace; see also Ref.\cite{Mintun2015} for an alternative proposal to evade the puzzles raised in Ref.\cite{Almheiri2014}. They also considered a set of states that are generated by applying bounded number of bulk local operators on some state with a fixed semi-classical background. Their proposal, which we refer to as the ADH proposal, is that the subspace spanned by these vectors can be interpreted as a quantum error correcting code in a certain sense.\footnote{Here we refrain from discussing Ref.\cite{Mintun2015}, as our tools in the present form do not seem to yield any nontrivial conclusions for it at the moment.}

The purpose of this paper is to point out a nontrivial implication of the ADH proposal: that certain operators lying oustide of the causal wedge can be reconstructed as well. A crucial notion in this context is the entanglement wedge;  Given a boundary subregion $A$, the entanglement wedge is formally defined as the domain of dependence of a bulk spacelike surface whose boundary is a union of $A$ and the minimal surface that is anchored on the boundary of $A$\cite{Czech2012,Wall2012,Headrick2014a,Jafferis2014}. The entanglement wedge is in general strictly larger than the causal wedge\cite{Headrick2014a}, and it was recently conjectured that operators that lie in it can be reconstructed from the corresponding boundary subregion; see also \cite{Czech2012,Wall2012}. This paper proves this conjecture within the proposal of Ref.\cite{Almheiri2014} for certain choices of boundary regions, modulo the caveats discussed below.

In order to explain our result, it will be instructive to first discuss an idealized limit, in which the AdS/Rindler reconstruction for operators that are sufficiently well-localized in the causal wedge works perfectly  and the Ryu-Takayanagi(RT) formula holds without any quantum corrections. In this limit,  entanglement wedge reconstruction works perfectly for certain choices of boundary subregions; see Fig.\ref{fig:Example} for an example. Of course, the reality is that we expect $1/N$ corrections to the AdS/Rindler reconstruction and quantum corrections to the RT formula\cite{Lewkowycz2013,Faulkner2013}. We carefully analyze  these correction terms and study the various limits in which their combined effects vanish, at which point we arrive at the same conclusion as before.

Our argument relies on the RT formula\cite{Ryu2006} and a few elementary facts in quantum information theory. We observe that the RT formula imposes a nontrivial structure on the underlying state, and then study the implication of this structure using quantum information theory.  In the rest of the paper, we review the proposal of Ref.\cite{Almheiri2014}, study the idealized limit in which the quantum correction vanishes, and then discuss the role of the correction terms.

During the completion of this manuscript, Ref.\cite{Dong2016} came to our attention, in which the authors addressed many of the same issues discussed herewithin. We include at the end of this manuscript a discussion of the similarities and differences between the two works.



\section{ADH Proposal}
In Ref.\cite{Almheiri2014}, the authors proposed to interpret  a certain subspace of a quantum gravity theory in pure AdS space as a quantum error correcting code. Here we review their proposal, making emphasis on the aspects that are pertinent to our analysis.

Our analysis begins by first declaring the existence of some state $\ket{\Omega}$ which has a good semi-classical description.  As we said, we shall choose the underlying geometry to be that of the pure AdS space. One can formally consider a set of operators $\phi_x$ at a bulk spacetime point $x$ and consider a subspace spanned by states of the form of $\phi_x \ket{\Omega}$. Any subspace is formally a quantum error correcting code, but of course this in itself is a vacuous statement; one needs to specify how the subspace is embedded into a Hilbert space for which we have a good handle on its structure. The purported isometric equivalence of quantum gravity theory and CFT implies that these states can be isometrically embedded into the Hilbert space of the CFT.

The main body of the ADH proposal concerns the nature of this subspace on the CFT side. For any state in this subspace, we can consider an auxiliary system $R$ that purifies the state. The proposal states that for any such state
\begin{equation}
\|\rho_{A^cR} - \rho_{A^c} \otimes \rho_R \|_1 \ll 1, \label{eq:ADH_proposal}
\end{equation}
where $\| \cdots \|_1$ is the trace norm, provided that the bulk point $x$ is sufficiently well-localized in the causal wedge of a boundary subregion $A$.\footnote{The actual proposal is more intricate, and even covers the operators that are near the causal wedge. We do not study such a case here.}

One immediate implication of the ADH proposal is the following:
\begin{equation}
\|\rho_{A^c} - \sigma_{A^c} \|_1 \ll 1,\label{eq:ADH_modified}
\end{equation}
for any two reduced density matrices in the subspace\footnote{Here the parameter that determines the smallness is smaller or equal to that of Eq.\ref{eq:ADH_proposal}.}; see Appendix \ref{section:appendix}. As we shall see in Section \ref{section:QC}, the above equation implies that the action of a bulk operator $\phi_x$ can be reproduced by an action of an operator on a boundary region $A$ if $x$ is in the causal wedge of $A$.\cite{Almheiri2014} Specifically, there exists some operator $\Phi_{A}$ on $A$ such that
\begin{equation}
V\phi_x \ket{\Omega} \approx \Phi_{A}V\ket{\Omega} \label{eq:reconstruction}
\end{equation}
 where $V$ is the isometry from the subspace of the gravity theory to the CFT Hilbert space.\footnote{The approximate equivalence implies that the two rays are nearly aligned.}

We must also consider the case where $x$ lies outside of the causal wedge of $A$. The so called entanglement wedge hypothesis asserts that even such bulk operators may be reconstructed from $A,$ provided that $x$ lies in the entanglement wedge of $A$.\cite{Headrick2014a} As we have already briefly discussed in the introduction, the entanglement wedge of a boundary region $A$ refers to the domain of dependence of a bulk spacelike surface whose boundary is a union of $A$ and the minimal surface that is anchored on the boundary of $A$; see Fig.\ref{fig:Example} for an example. The entanglement wedge is in general larger than the causal wedge, and as such, the entanglement wedge hypothesis demands much more than Eq.\ref{eq:ADH_proposal} for it to be true.

Now we can formally address our main claim. Assuming the RT formula holds, Eq.\ref{eq:reconstruction} holds if $x$ lies in the entanglement wedge of $A$ for the choice of boundary regions depicted in Fig.\ref{fig:Example}.\footnote{Our argument is actually more general, but perhaps not general enough to ensure the same conclusion for all possible boundary subregions.} The approximate nature of this claim at the moment is hiding a number of subtleties. We shall address these issues after we study an idealized limit in which every correction term vanishes.

\section{Idealized limit}
By idealized limit we mean three things. First, the right hand side of Eq.\ref{eq:ADH_proposal} is  0. By the analysis delineated in the Appendix, this implies that the right hand side of Eq.\ref{eq:ADH_modified} is 0 a well.  Second, the RT formula holds exactly. That is, for a CFT that is dual to a well-defined semiclassical gravity, the entanglement entropy of a CFT subregion $A$ is
\begin{equation}
S(A) = \frac{\text{Area}(\gamma_A)}{4G},\label{eq:RT_formula}
\end{equation}
where $\gamma_A$ is a static minimal surface in AdS space whose boundary is given by $\partial A$. Third, $G\to 0$ so that the gravitational backreaction of the bulk fields are nonexistent. Let us first begin with a simple example that demonstrates our argument.
\subsection{Simple example}
\begin{figure}[h]
\begin{center}
\subfigure[Causal surface]{
\begin{tikzpicture}[scale=0.6]
\draw[black] (0,0) circle (3);
\draw[red] plot [smooth, tension=1.3] coordinates {(45:3) (90:1.5) (135:3) };
\draw[red] plot [smooth, tension=1.3] coordinates {(45:3) (0:1.5) (-45:3) };
\draw[red] plot [smooth, tension=1.3] coordinates {(225:3) (-90:1.5) (-45:3) };
\draw[red] plot [smooth, tension=1.3] coordinates {(225:3) (180:1.5) (135:3) };
\node[below] at (-90:3) {$A$};
\node[above] at (90:3) {$A$};
\node[right] at (0:3) {$B$};
\node[left] at (180:3) {$B$};
\node[above] at (0,0) {$\phi_x$};
\filldraw[black] (0,0) circle (0.1);
\end{tikzpicture}
}
\subfigure[Entangling surface]{
\begin{tikzpicture}[scale=0.6]
\draw[] (0,0) circle (3);
\draw[blue] plot [smooth, tension=1.3] coordinates {(45:3) (0:1.5) (-45:3) };
\draw[blue] plot [smooth, tension=1.3] coordinates {(225:3) (180:1.5) (135:3) };
\node[below] at (-90:3) {$A$};
\node[above] at (90:3) {$A$};
\node[right] at (0:3) {$B$};
\node[left] at (180:3) {$B$};
\node[above] at (0,0) {$\phi_x$};
\filldraw[black] (0,0) circle (0.1);
\end{tikzpicture}
}
\end{center}
\caption{A single-time slice of the causal wedge of two regions $A$ and $B.$ It is assumed that $|A|>|B|$. The causal/entanglement wedge is enclosed by a boundary subregion and its corresponding causal(red)/entanging(blue) surface.\label{fig:Example}}
\end{figure}
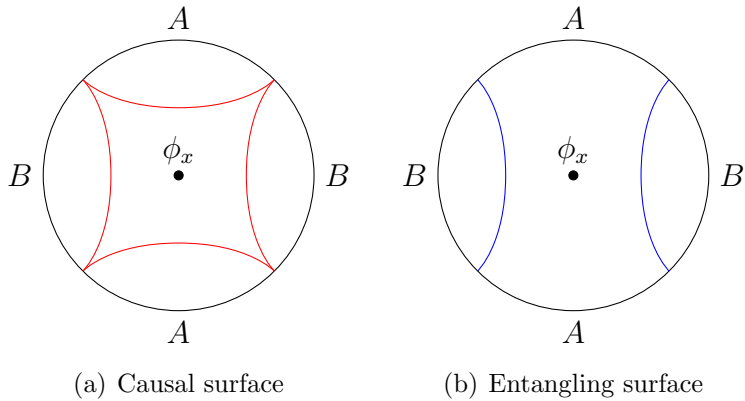

There is a paradigmatic example that appears in the discussion of causal and entanglement wedges; see Fig.\ref{fig:Example}. The interior of the circle represents the bulk of the AdS, and the boundary represents the boundary CFT, both at a single-time slice. The size of $A$ and $B$ are comparable, but we do assume that ths size of $A$ is larger than that of $B$.

For $B,$ the entanglement and the boundary wedge coincides. Therefore, for operators that lie in the entanglement wedge of $B,$ the entanglement wedge conjecture immediately follows. A nontrivial question concerns operators  that lie inside the entanglement wedge of $A,$ but outside of the causal wedge of $A.$

Without loss of generality, consider the two-dimensional subspace spanned by $V\ket{\Omega}$ and $V\phi_x \ket{\Omega}$ where $x$ is well-localized in the entanglement wedge of $A.$ For any states in this subspace, their reduced density matrices on $B$ are identical. To see this, let us denote each connected components of $B$ as $B_1$ and $B_2$. The operator $\phi_x$ lies in the causal wedge of the complement of $B_1$, as well as the causal wedge of the complement of $B_2$. As such, the ADH proposal implies that the reduced density matrices of any states in this subspace are identical, both on region $B_1$ and $B_2$; see Eq.\ref{eq:ADH_modified}.

Does this imply that the reduced density matrices are identical on the union of $B_1$ and $B_2?$ The answer is yes, provided that the RT formula holds. Specifically, the formula implies that the mutual information between $B_1$ and $B_2$ is $0:$
\begin{equation}
I(B_1:B_2)=S(B_1)+S(B_2) - S(B_1B_2) = 0,
\end{equation}
provided that the sum of the area of the minimal  surfaces of $B_1$ and $B_2$ are equal to area of the minimal surface of $B$.\footnote{It is reasonable to expect this to happen in the $G\to 0$ limit, as the gravitational backreaction to the underlying geometry will be negligible.}
This means that $B_1$ and $B_2$ are completely uncorrelated. Their joint reduced density matrices factorize. Both $V\ket{\Omega}$ and $V\phi_x \ket{\Omega}$ have a reduced density matrices of a form $\rho_{B_1} \otimes \rho_{B_2}$ on $B$.  Since the two states have identical density matrices on both $B_1$ and $B_2$, their reduced density matrices on the entire $B$ is identical as well.

The fact that the two states have identical density matrices on $B$ implies that there is an operator on $A=B^c$ that maps one state to another. Both states, after normalization, have the following Schmidt decomposition:
\begin{equation}
\sum_n \sqrt{p_n} \ket{n}_A \ket{n}_B,\label{eq:Schmidt}
\end{equation}
where $p_n$ and $\ket{n}_B$ are the eigenvalues and eigenstates of the reduced density matrix of $B$. The only possible difference between the two states is the choice of the basis states $\{ \ket{n}_A\}$. As any set of basis states can be mapped into each other by some unitary operator, we can conclude that there exists some operator on $A$ such that it maps $V\ket{\Omega}$ into $V\phi_x\ket{\Omega}$.

\subsection{Extensions and limitations}
An implicitly used fact in the previous analysis is that the causal wedge and entanglement wedge coincides for a single boundary interval. This is a simple corollary of a more general fact: that both wedges coincide for a spherical boundary region and its complement\cite{Hubeny2012}. Let us again assume that a bulk operator $\phi_x$ is located at a bulk point $x$ in the pure AdS space, and also that it is located inside the entanglement wedge of $A$. It is clear that the previous analysis carries over, provided that $B=A^c$ is a union of a disjoint spherical regions that are sufficiently well-separated from each other so as to ensure that the multipartite mutual information vanishes:
\begin{align}
I(B_1:B_2:\cdots B_n) &= [\sum_{i=1}^{n}S(B_i)] - S(B) \nonumber \\
&=0. \label{eq:Multipartite_Mi}
\end{align}

Indeed, for such choice of $B$, the causal wedge of the complement of $B_i$ includes $x$ for all $i$. The ADH proposal thus implies that  $\ket{\Omega}$ and $\phi_x \ket{\Omega}$ are identical on $B_i$ for all $i$. Now observe that the reduced density matrix factorizes over $B_i$ due to Eq.\ref{eq:Multipartite_Mi}. Therefore, $\ket{\Omega}$ and $\phi_x\ket{\Omega}$ have the same  density matrix over $B$. Both states have the Schmidt decomposition of the form of Eq.\ref{eq:Schmidt}. There thus exists an operator $A$ that reproduces the action of $\phi_x$.

Unfortunately our argument in its present form has only limited applicability for general asymptotically AdS space. The fact that the entanglement and the causal wedge coincides for a spherical region and its complement is a fact that is only applicable to the pure AdS space; otherwise, the entanglement wedge in general lies strictly beyond the causal wedge\cite{Hubeny2012}.

\section{Quantum corrections\label{section:QC}}
There are two subtleties that we glossed over so far.  First, the right hand side of Eq.\ref{eq:ADH_proposal} will not be zero in general. Second, there is also a known correction term to the RT formula. Both of these effects can be handled, as we discuss below.

It will be convenient to designate two parameters, each of which quantify these correction terms. To be concrete, let us denote the upper bound on the trace distance of Eq.\ref{eq:ADH_proposal} to be $\epsilon.$ The second parameter quantifies the correction to the holographic mutual information, not the entanglement entropy. Specifically, we assume that the multipartite mutual information(Eq.\ref{eq:Multipartite_Mi}) is $\delta$. Then the previous analysis carries over, with a correction term that scales as $O(\delta^{\frac{1}{2}} + n\epsilon)$, where $n$ is the number of connected components on $B$.

As before, we can first establish the closenss of the reduced density matrices for $V\ket{\Omega}$ and $V\phi_x \ket{\Omega}$ on some region.
To see this, first recall the Pinsker's inequality:
\begin{equation}
\frac{1}{2} \|\rho - \sigma \|_1^2\leq D(\rho \| \sigma),
\end{equation}
where $D(\rho \| \sigma) = \Tr(\rho (\log \rho - \log \sigma))$ is the relative entropy between $\rho$ and $\sigma$\cite{Wilde2011}. The multipartite mutual information can be rewritten as a relative entropy between $\rho_{B}$ and $\rho_{B_1} \otimes \rho_{B_2} \otimes \cdots \rho_{B_n}$. As such, these two states are close to each other in trace norm up to a $O(\delta^{\frac{1}{2}})$ correction. Let us denote the reduced density matrices of $V\ket{\Omega}$ as $\rho_{B_1}, \rho_{B_2}, \cdots$ and the reduced density matrices of $V\phi_x\ket{\Omega}$ as $\rho_{B_1}', \rho_{B_2}', \cdots$. By our assumption $\|\rho_{B_i} - \rho_{B_i}'' \|_1\leq \epsilon$. By recursively applying the logic of the following form,
\begin{align}
\|\rho_{B_1}\otimes \rho_{B_2} - \rho_{B_1}' \otimes \rho_{B_2}' \|_1 &\leq \|\rho_{B_1}\otimes \rho_{B_2} - \rho_{B_1} \otimes \rho_{B_2}' \|_1 + \|\rho_{B_1}\otimes \rho_{B_2'} - \rho_{B_1}' \otimes \rho_{B_2}' \|_1 \nonumber \\
&\leq \|\rho_{B_2} - \rho_{B_2}' \|_1 + \|\rho_{B_1} - \rho_{B_1}'\|_1,
\end{align}
one can conclude that $\|\rho_{B_1}\otimes \cdots \otimes \rho_{B_n}  - \rho_{B_1}'\otimes \cdots \otimes \rho_{B_n}'\|_1 \leq n\epsilon.$ Pinsker's inequality implies that $|\rho_{B_1}\otimes \cdots \otimes \rho_{B_n}$($|\rho_{B_1}'\otimes \cdots \otimes \rho_{B_n}'$) is close to $\rho_B$($\rho_B'$) up to a trace distance of $O(\delta^{\frac{1}{2}})$. Now using triangle inequality for the trace norm, we conclude that $\rho_B$ and $\rho_B'$ is close to each other with a trace distance of $O(\delta^{\frac{1}{2}} + n\epsilon)$.

A generalization of the observation that led to Eq.\ref{eq:Schmidt} is known as Uhlmann's theorem\cite{Uhlmann1976}. The theorem states that the fidelity between two reduced density matrices, is equal to the maximum overlap between the purification of the two states:
\begin{align}
F(\rho, \sigma) &:= \| \rho^{\frac{1}{2}} \sigma^{\frac{1}{2}}\|_1 \\
&= \sup_{U_R} \bra{\psi_{\rho}} U_R\ket{\psi_{\sigma}},
\end{align}
where $\ket{\psi_{\rho}}$($\ket{\psi_{\sigma}}$) is a purification of $\rho$($\sigma$) and $U_R$ is a unitary operator acting on the purifying space. It is a well-known fact that the fidelity between two states is close to $1$ if and only if their trace distance is small\cite{Nielsen2000}:
\begin{equation}
  1- \frac{1}{2}\|\rho - \sigma \|_1  \leq F(\rho, \sigma).
\end{equation}

Note that both $V\ket{\Omega}$ and $V\phi_x\ket{\Omega}$ are pure states, and as such, the purifying space of their reduced density matrices on $B$ is $A.$ Plugging in the derived expression for the distance between $\rho$ and $\sigma$, we arrive at the main result. Namely, there exists an operator on $A$ such that Eq.\ref{eq:reconstruction} holds, up to a correction of $O(n\epsilon + \delta^{\frac{1}{2}})$.

It is interesting to study the various limits in which this correction term becomes small. In the limit that $B$ consists of disjoint regions with a distance scale much larger than the lengthscale of each regions, it is reasonable to expect the multipartite mutual information to be determined by the following asymptotic form:
\begin{equation}
I(B_1:B_2:\cdots :B_n) \sim \frac{C_{\Delta}}{r^{\Delta}}.
\end{equation},
where $r$ is the shortest distance scale, $\Delta$ is the lowest scaling dimension and $C_{\Delta}$ is the corresponding OPE coefficient.

The bipartite mutual information in (1+1)D CFT can be obtained by evaluating the expectation values of the twist operators for $n-$sheeted Riemann surface, and then analytically continuing the result to $n\to 1$\cite{Calabrese2004}. Since the twist operator is primary, the result should only depend on the cross ratio between the four regions depicted in Fig.\ref{fig:Example}. Specifically, setting the lengths of the two intervals of $A$ as $x_1,x_2$ and the lengths of the two intervals as $x_3,x_4$, the mutual information should be a function of $x=\frac{x_1x_2}{x_3x_4}$. Its dependence on $x$ should scale as $x^{\Delta}$, and as such, in $\Delta \to \infty$ limit it is reasonable to expect it to vanish for $0<x<1.$ A missing part of the argument concerns the behavior of the OPE coefficient in this limit. As long as it grows subexponentially with $\Delta$, our claim should remain intact. Intriguingly, extremal CFTs with  large central charge by definition should have this property\cite{Witten2007}. Our work does not seem to shed any new light on the existence of such CFTs. However, it does strongly suggest that such a theory, if it is dual to the pure quantum gravity theory in (2+1)D AdS as conjectured in Ref.\cite{Witten2007}, should admit entanglement wedge reconstruction.

Unfortunately, this line of  reasoning is more limited to the well-studied instances of the AdS/CFT with extra compactified dimension. The mutual information will always contain a term of the form of $\frac{1}{r^{\Delta}}$, and the existence of the Kaluza-Klein excitations necessarily puts a nontrivial upper bound on $\Delta$ according to the standard dictionary\cite{Gubser1998,Witten1998}.

\section{Discussion}
The ADH proposal, whilst being a general framework that is independent of the details of the AdS/CFT correspondence, already puts a strong constraint on the precursor problem. The reason is that the states appearing in this context have a special universal structure that is inherited from the RT formula. Indeed, we were able to say something nontrivial about bulk operators that lie beyond the causal wedge by incorporating this fact.

Modulo our assumptions, our work constitute as a partial proof of the entanglement wedge conjecture. Of course, this is not without any deficiencies. Most importantly, the argument is based on the ADH proposal, and we did not justify this proposal from more fundamental principles. Also, our argument cannot be applied to operators that are located close to the causal/entangling surface, as such operators cannot be reconstructed from the boundary region in a standard quantum error correcting sense\cite{Almheiri2014}. It is also worth noting that causal wedge and entanglement wedge differ from each other in general in asymptotically AdS space\cite{Hubeny2012}, even for spherical boundary regions. A different line of reasoning seems necessary in order to study these cases.

In the language of quantum error correcting code, our analysis can be interpreted as enlarging a correctable region of the code from the assumptions about the codewords. The assumption used here is the RT formula, but an insight in a similar vein has been already commonly used in deriving fundamental tradeoff bounds for quantum error correcting codes.\cite{Bravyi2009} It turns out that reformulating the problem this way and generalizing these insights leads to more fundamental constraints, as it shall be demonstrated elsewhere.


We note that the work of Dong, Harlow, and Wall\cite{Dong2016} also makes significant progress on the causal wedge/entanglement wedge question. There are two key differences between their proposal and this one, however.

The first difference is the degree to which Ref.\cite{Jafferis2015} is used in the two constructions; while it is central to the DHW construction, our construction does not make direct use of it. This is perhaps related to the fact that their analysis applies to the entirety of asymptotically AdS spacetimes, while ours hold more specifically to small deformations around vacuum AdS.

The second difference is that their construction requires that for two states in the bulk localized in the complement of the entanglement wedge are indistinguishable from each other, with relative entropy 0. While this seems reasonable, we are not sure how to prove this claim in full generality, and thus do not use it in our construction.

\bigskip
\centerline{\bf{Acknowledgements}}
We would like to thank Xi Dong for useful discussions.
We thank the Perimeter Institute for hospitality during parts of this project.  Ning Bao is supported by the DOE Grant DE-SC0011632 and funding provided by the Institute for Quantum Information and Matter, an NSF Physics Frontiers Center (NSF Grant PHY-1125565) with support of the Gordon and Betty Moore Foundation (GBMF-2644). He is also supported by the Walter Burke Institute of Theoretical Physics. Isaac Kim's research at Perimeter Institute is supported in part by the Government of Canada through NSERC and by the Province of Ontario through MRI.

\bibliographystyle{JHEP}
\bibliography{bib.bib}

\appendix

\section{Proof of Eq.\ref{eq:ADH_modified}\label{section:appendix}}
Here we explain why Eq.\ref{eq:ADH_modified} follows from Eq.\ref{eq:ADH_proposal}. Our explanation is based on a few well-known but nontrivial facts. As before, let us use $V$ to formally denote an isometry from the Hilbert space of the quantum gravity theory to the CFT Hilbert space. It is well-known that an isometry followed by a partial trace is precisely the linear map that is known as completely-positive trace preserving map\cite{Nielsen2000}. Here we choose the partial trace to be over $A.$

It is well-known that under Eq.\ref{eq:ADH_proposal} a region $A^c$ is approximately correctable\cite{Schumacher2007}. For an erasure over $A^c$, there exists a completely-postivie trace-preserving map $\mathcal{R}$ such that
\begin{equation}
\|\mathcal{R} \circ \mathcal{N}_{A}(\cdot) - \mathcal{I} \|_{\diamond} \ll 1,
\end{equation}
where $\| \cdots \|_{\diamond}$ is a diamond norm for superoperators\cite{Kitaev2002book}, $\mathcal{N}_A=\Tr_{A^c}(V(\cdot)V^{\dagger})$ and $\mathcal{I}$ is the identity superoperator.\footnote{Diamond norm is also known as cb-norm, an acronym of completely bounded norm.}

By the information-disturbance tradeoff relation\cite{Kretschmann2007}, the above inequality is satisfied if and only if
\begin{equation}
\|\mathcal{N}_{A^c} - \Phi_{A^c}\|_{\diamond} \ll 1,
\end{equation}
for some completely-positive trace preserving map $\Phi_{A^c}$ which outputs a fixed density matrix irrespective of the input. It is a standard fact that $\|\cdot \|_1 \leq \|\cdot \|_{\diamond}$ for any superoperators; see Ref.\cite{Kitaev2002book}. This proves Eq.\ref{eq:ADH_modified}.

\end{document}